\documentclass[aps,pra,twocolumn,groupedaddress,nofootinbib]{revtex4-1}

\pdfoutput=1

\usepackage{graphics}
\usepackage{epsfig}
\usepackage{color}
\usepackage{amsmath}
\usepackage{bm}
\usepackage[toc,page]{appendix}
\usepackage[hidelinks]{hyperref}
\usepackage[caption = false]{subfig}

\begin{document}




\title{Intensity dependence of Rydberg states} 

\author{L. Ortmann$^{1}$}
\email[]{ortmann@pks.mpg.de}
\author{C. Hofmann$^{1}$}
\author{A. S. Landsman$^{1,2}$}
\email[]{landsman@pks.mpg.de}
\affiliation{$^1$Max Planck Institute for the Physics of Complex Systems, N\"othnitzer Stra{\ss}e 38, D-01187 Dresden, Germany}
\affiliation{$^2$Department of Physics, Max Planck Postech, Pohang, Gyeongbuk 37673, Republic of Korea}

\date{\today}

\begin{abstract}
We investigate numerically and analytically the intensity dependence of the fraction of electrons that end up in a Rydberg state after strong-field ionization with linearly polarized light. We find that including the intensity dependent distribution of ionization times and non-adiabatic effects leads to a better understanding of experimental results.  Furthermore, we observe using Classical Trajectory Monte Carlo simulations that the intensity dependence of the Rydberg yield changes with wavelength and that the previously observed power-law dependence breaks down at longer wavelengths.  Our work suggests that Rydberg yield measurements can be used as an independent test for non-adiabaticity in strong field ionization.  
\end{abstract}

\maketitle

The liberation of the electron in the process of strong field ionization via tunneling \cite{keldysh1965ionization,corkum1993plasma,ivanov2005anatomy} does not necessarily lead to the electron leaving the atom for good \cite{nubbemeyer2008strong,shvetsov2009capture}. This effect that is often referred to as `frustrated tunneling ionization' (FTI) is understood by the low kinetic energy of some electrons at the end of the laser pulse which does not allow them to leave the Coulomb potential but results in their capture in a Rydberg state. 

This process is not only interesting because it produces neutral excited states, which are found to be useful tools in the investigation of other strong field effects \cite{eichmann2009acceleration,eilzer2014steering}, but it also leads to a better understanding of post-ionization dynamics \cite{eichmann2009acceleration,eilzer2014steering,zimmermann2018limit}. 

Even though the detection of neutral excited states poses some difficulties \cite{nubbemeyer2008strong}, the fact that about 10\% of the liberated electrons end up in a Rydberg state for typical strong field parameters makes it a process that needs to be taken into account in the investigation of many strong field effects \cite{manschwetus2009strong,li2014rescattering,lv2016comparative,liu2012low}. The fraction of electrons that are tunnel ionized and which end up in a Rydberg state was found to depend significantly on parameters of the laser field and the atomic potential, the experimental and theoretical investigation of which helped understand the underlying process of FTI better \cite{nubbemeyer2008strong,shvetsov2009capture,li2014rydberg,Eichmann2016}. 

In the present work, we focus on the intensity dependence of the ratio of tunnel-ionized electrons which end up in a Rydberg state when using linearly polarized light.  This observable has been previously measured by Nubbemeyer et al in \cite{nubbemeyer2008strong}. In \cite{shvetsov2009capture}, Shvetsov-Shilovski et al. have presented analytical estimations and numerical calculations for this experimental data.  Here, we build on this work by including non-adiabatic effects, as well as introducing further corrections and expansions of the theory.   We find an analytical dependence of Rydberg yield on intensity that agrees better with the experimental results in \cite{nubbemeyer2008strong}.  Additionally, we describe wavelength dependent effects, which to the best of our knowledge, have not been predicted so far and should be experimentally measurable.

The insights gained in the present study are not only restricted to Rydberg states but address the more general questions of which approximations are useful to describe (i) the initial conditions at the tunnel exit and (ii) the movement of the electron in the superposed potential of the laser and the parent ion.   These approximations are the basis of many classical trajectory methods \cite{yudin2001nonadiabatic,shvetsov2016semiclassical}, and are fundamental to our interpretation of many high profile experiments, including recent attoclock measurements \cite{landsman14,camus2017}.   The present work therefore demonstrates in what way Rydberg atoms can be used to give answers to these questions and to thus track the electron motion in a strong field ionization process.  

In particular, our results provide support for the importance of non-adiabatic effects in strong field ionization -- a much debated question that has previously been addressed by investigating photoelectron momenta distributions \cite{boge2013probing,hofmann2016non,arissian10}.  These investigations, however, have proved to be inconclusive, with some experiments confirming adiabatic assumptions \cite{boge2013probing,arissian10}, while others pointing to relevance of non-adiabatic effects under typical strong field ionization conditions \cite{hofmann2016non,nirit15}.  

Since Rydberg yield is measured under different experimental conditions and represents a different class of electrons (inaccessible in typical strong field experiments), its experimental measurements provide an independent test of the prominence of non-adiabatic effects in strong field ionization.  Furthermore, this non-adiabaticity manifests itself in the power-law dependence as a function of intensity.  Since the absolute value of intensity is therefore not important, the results do not depend on the calibration procedure (something that has been a serious issue in prior studies \cite{boge2013probing,hofmann2016non,arissian10}).


Even though there are some effects in FTI that can only be understood based on the time-dependent Schr\"odinger equation \cite{popruzhenko2017quantum,lv2016comparative}, it has been found that electrons that end up in a Rydberg state can be described very well in a semiclassical approximation \cite{shvetsov2009capture,huang2013survival,zhang2014generation,xiong2016correspondence,landsman2013rydberg}. One semiclassical method that is widely used and that also we will use in this paper is called the Classical Trajectory Monte Carlo (CTMC) method \cite{rose1997ultrafast,cohen2001reexamination,landsman04,comtois2005observation}. In this framework, the electron is born at the tunnel exit at a time $t_0$ with an initial velocity $v_{\perp,0}$ perpendicular to the polarization direction where $t_0$ and $v_{\perp,0}$ are sampled according to a probability distribution. Each electron is then propagated in the superposed laser and atomic field solving Newton's equations. 
In order to determine which electron is captured in a Rydberg state, we evaluate the total electron energy at a time $\tau$ when the pulse has passed. The final total energy $E$ has to be negative in the case of FTI:
\begin{equation}
 E = \frac{v^2}{2}-\frac{1}{r} < 0\label{eq:Rydberg-condition}.
\end{equation}
Atomic units are used throughout the paper, unless otherwise specified.\\

\begin{figure}
  \centering
  \includegraphics[width=0.45\textwidth]{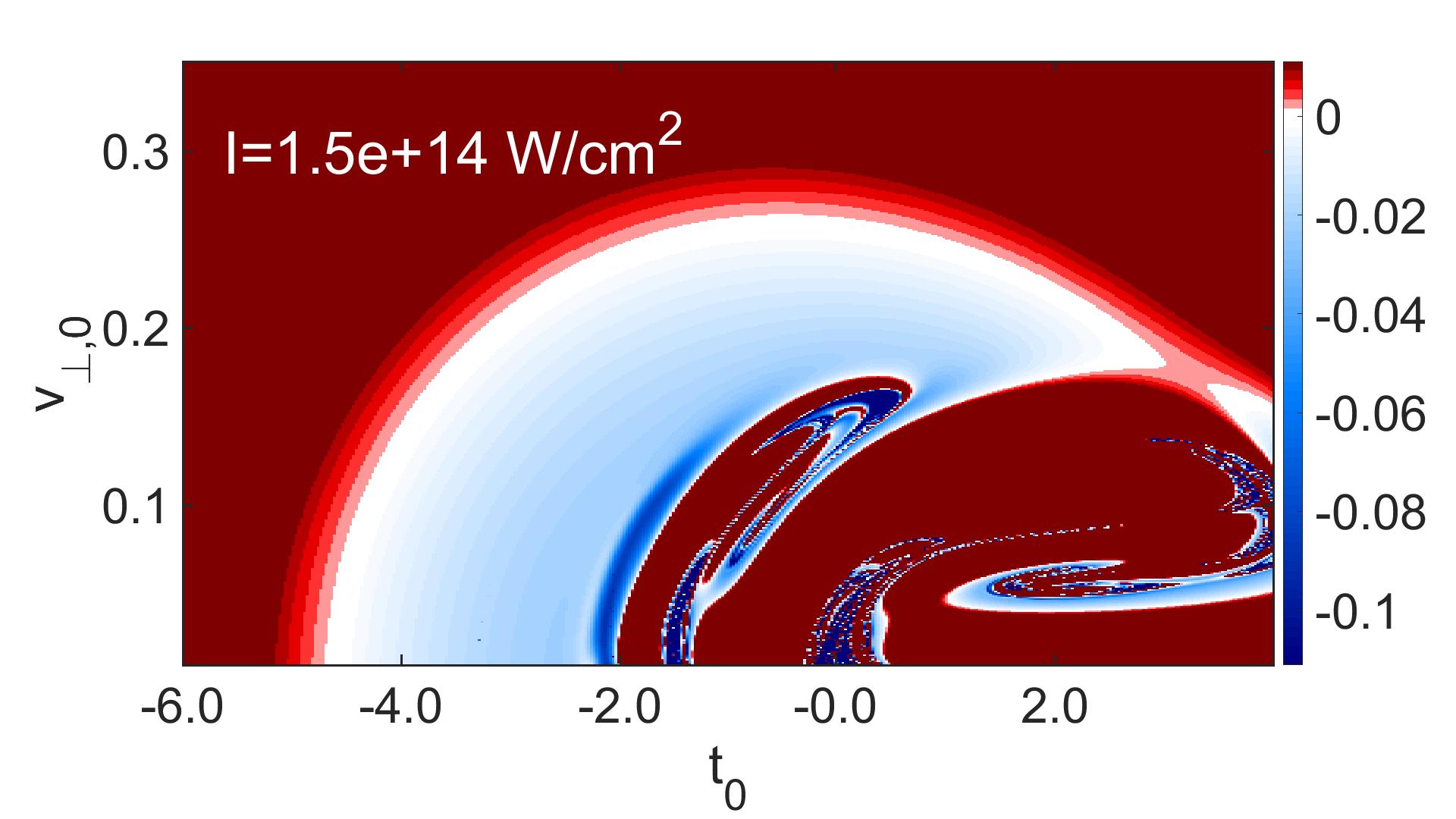}
  \includegraphics[width=0.45\textwidth]{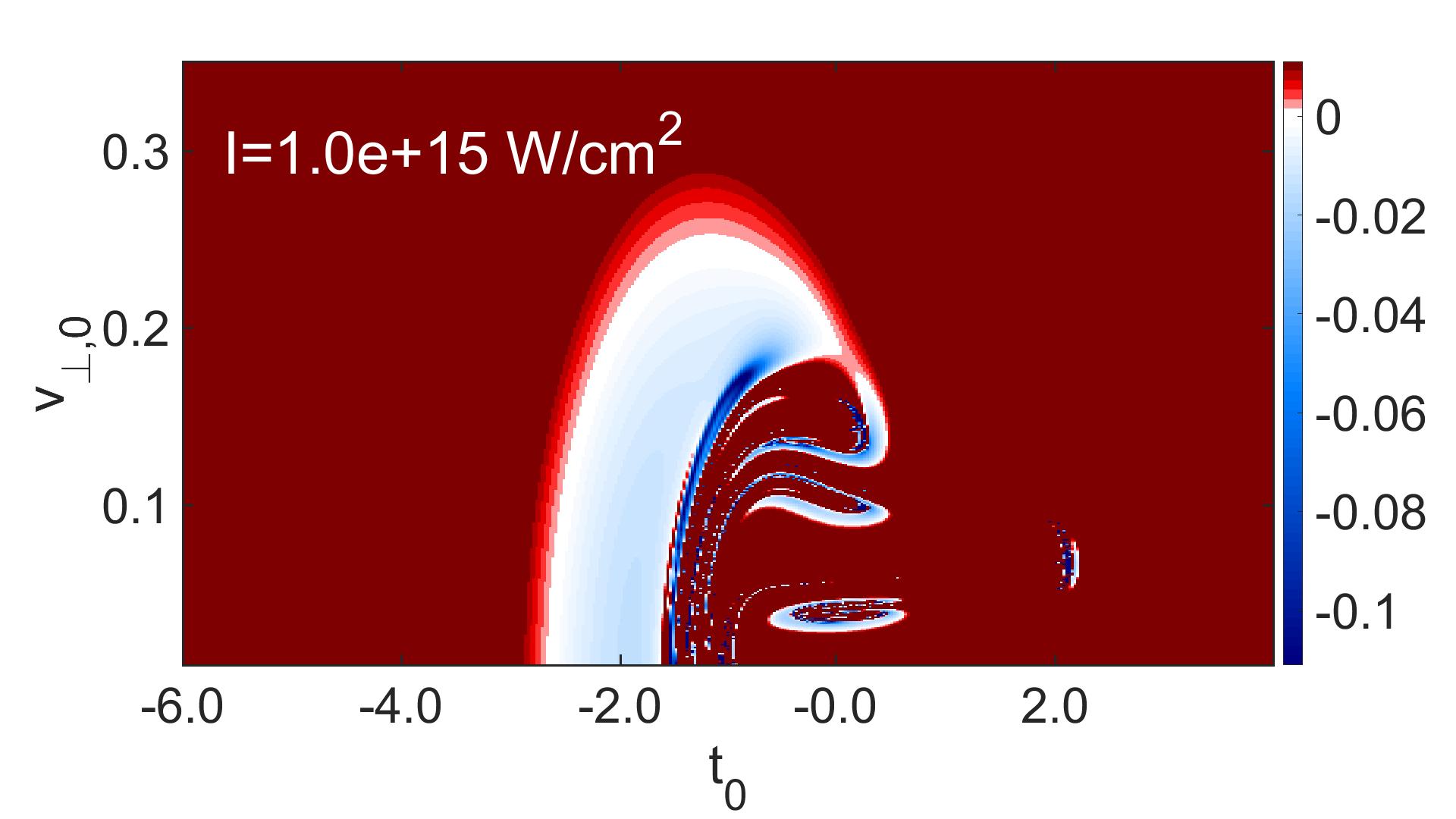}
  \caption{The total energy (see colorbar) at the end of the pulse for electrons ionized at $t_0$ with initial transverse momentum $v_{\perp,0}$. All quantities are given in atomic units. The laser pulse here was chosen to have a wavelength of $\lambda=800 \, \mathrm{nm}$ and 8 cycles for two different intensities $I$ specified in the plots. Rydberg states have a negative total energy and one can see how the Rydberg area shrinks for larger intensities.}
  \label{fig:Rydberg-ellipse}
\end{figure}

We define the Rydberg yield as the ratio of the number $N^*$ of electrons which are captured in a Rydberg state to the number $N$ of all electrons which tunneled through the potential barrier.
As is the case in \cite{shvetsov2009capture}, we initially assume a constant distribution of ionization phases $\phi=\omega t_0$ and initial transverse velocities $v_{\perp,0}$ in the $\phi$ - $v_{\perp,0}$-plane,
meaning the Rydberg yield is estimated to be proportional to the ratio $\Sigma^*/\Sigma$ of the areas $\Sigma^*$ and $\Sigma$ which are obtained by integrating in the $\phi$ - $v_{\perp,0}$-plane over the regime of the Rydberg or ionization events, respectively. 
Fig. \ref{fig:Rydberg-ellipse} displays the Rydberg area for two different intensities for ionization during the central half-cycle.
The estimate for the area $\Sigma^*$ of Rydberg states in \cite{shvetsov2009capture} is derived for ionization in that central half-cycle giving
\begin{equation}
  \Sigma^* \propto \frac{\omega}{F_0 \tau^{3/2}} \left(1- 2 \frac{F_0}{(2 I_p)^2}  \right)^{-1}\label{eq:Sigma*-Shvetsov}, 
\end{equation}
where $F_0$ denotes the maximal field strength and $I_p$ the ionization potential. 
Furthermore, in \cite{shvetsov2009capture} the area $\Sigma$ is assumed to be proportional to the width $\sigma_{v_\perp}$ of the 
distribution of the initial transverse velocity $v_{\perp,0}$ as described by \cite{delone1991energy, ammosov1986tunnel} with
\begin{equation}
 \Sigma \propto \sigma_{v_\perp} \propto \sqrt{F_0},
\end{equation}
where the relation $\sigma_{v_\perp} \propto \sqrt{F_0}$ is not trivial and is discussed in more detail in Appendix \ref{app:AppendixA}.
Thus, the Rydberg yield is estimated to be proportional to 
\begin{equation}
  N^*/N \propto \frac{\omega}{F_0^{3/2} \tau^{3/2}} \left(1- 2 \frac{F_0}{(2 I_p)^2}  \right)^{-1}  \label{eq:power-law} \\
\end{equation}
where the last factor can be neglected for $2F_0 \ll (2I_p)^2 $. Setting all parameters except the intensity $I$ to a constant we thus arrive at the power law $N^*/N \propto F_0^{-3/2} \propto I^{-0.75}$, which is the result presented in \cite{shvetsov2009capture}.

\begin{figure}
  \centering
  \includegraphics[width=0.45\textwidth]{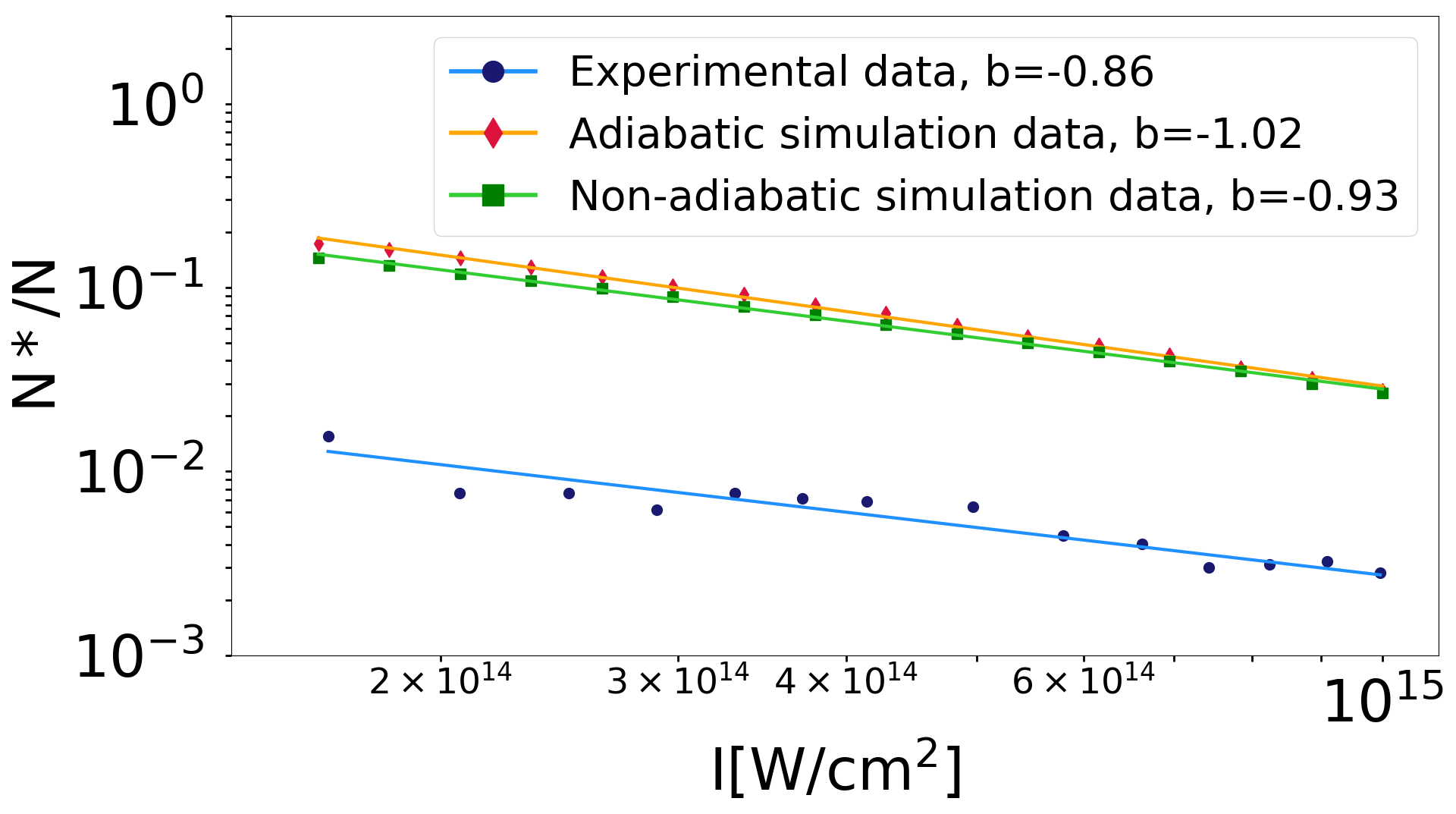}
  \caption{Rydberg yield for the parameters found in \cite{nubbemeyer2008strong}: $I=1.4\cdot 10^{14}-10^{15}$, FWHM of pulse envelope $= 30$ fs, $\lambda=800$ nm, He atom with $I_p=0.9$. The experimental yield (blue dot) was extracted from \cite{nubbemeyer2008strong}. The adiabatic CTMC simulation (red diamond) was done using the ADK distribution \cite{delone1991energy, ammosov1986tunnel} and the non-adiabatic simulation (green square) is based on \cite{mur2001energy}. The power law used for fitting is described by $N^*/N = a \cdot I^b$ with $b$ given in the legend. The fitting results are represented by lines. Note that the lower absolute values of the experimental yields are due to the decay of the excited states which is not accounted for here (for details see \cite{nubbemeyer2008strong}). As we expect the decay rate to be the same over the depicted intensity regime, this should not affect the decline though.}
  \label{New-exponent-for-Shvetsov}
\end{figure}

However, also the width $\sigma_{\phi}$ of the ionization phase depends on the laser intensity and we should take account of that. As shown in Appendix \ref{app:AppendixA} and as often used \cite{popov2004tunnel,ortmann2018analysis} the adiabatic ADK distribution for ionizations phases $\phi=w\cdot t$ \cite{delone1991energy, ammosov1986tunnel}
\begin{equation}
 P(\phi) \propto \exp \left(-\frac{2(2 I_p(\phi))^{3/2}}{3 F_0 \cdot |\cos(\phi)|}\right) \label{eq:ADK-probability}
\end{equation}
can be approximated as a Gaussian function with an intensity dependent width $\sigma_{\phi}$ that can be estimated as being proportional to $\sqrt{F_0}$. Consequently, we should set $N \propto \sigma_{v_\perp} \cdot \sigma_{\phi} \propto \sqrt{F_0} \cdot \sqrt{F_0} = F_0 = \sqrt{I}$ obtaining $N^*/N \propto I^{-1}$. This conclusion enables a better understanding of the adiabatic CTMC simulation results displayed in Fig. \ref{New-exponent-for-Shvetsov} where a power law fit to the data yields an exponent of $-1.02$.

From the experiment reported in \cite{nubbemeyer2008strong}, the ratio $N^*/N$ can be extracted for various intensities. These values show an intensity dependence of 
\begin{equation}
 N^*/N \propto I^{-0.86},
\end{equation}
displayed as blue line in Fig. \ref{New-exponent-for-Shvetsov}.
So, even though taking into account the intensity-dependent phase-width in the analytical estimation, which shifted the power law exponent from $b=-0.75$ as obtained in \cite{shvetsov2009capture} to $b=-1$, was well captured by the adiabatic CTMC simulations giving $b=-1.02$, we still do not fully understand the experimental result of $b=-0.86$ in this framework.
However, when looking at the adiabaticity parameter $\gamma=\omega \sqrt{2 I_p}/F$ \cite{keldysh1965ionization}, we find that, for the intensity regime of $I=1.4\cdot 10^{14}-10^{15} \, \mathrm{W/cm^2}$ at $\lambda=800 \, \mathrm{nm}$, $\gamma$ ranges from 0.5 to 1.2.  This is the typical strong field ionization regime, where the relevance of non-adiabatic effects is under debate \cite{boge2013probing,hofmann2016non,arissian10}.  

We now show that non-adiabatic effects can be observed in Rydberg yield measurements from the power-law dependence alone.  This eliminates the concerns about intensity calibration that has haunted prior experiments attempting to observe non-adiabatic effects by measuring electron momenta distributions \cite{boge2013probing,hofmann2016non}.

In Fig. \ref{New-exponent-for-Shvetsov}, CTMC simulation results are depicted in green (squares) where the non-adiabatic PPT ionization probability described in \cite{mur2001energy} and \cite{perelomov1966ionization} was used to generate the initial conditions. For a detailed description of this simulation see \cite{hofmann2014interpreting}. A power law fit to this data yields $N^*/N \propto I^{-0.93}$, which improves the CTMC prediction and gives the closest quantitative agreement with the experimental value of $b=-0.86$ of all discussed models.
 
\begin{figure}
  \centering
  \includegraphics[width=0.45\textwidth]{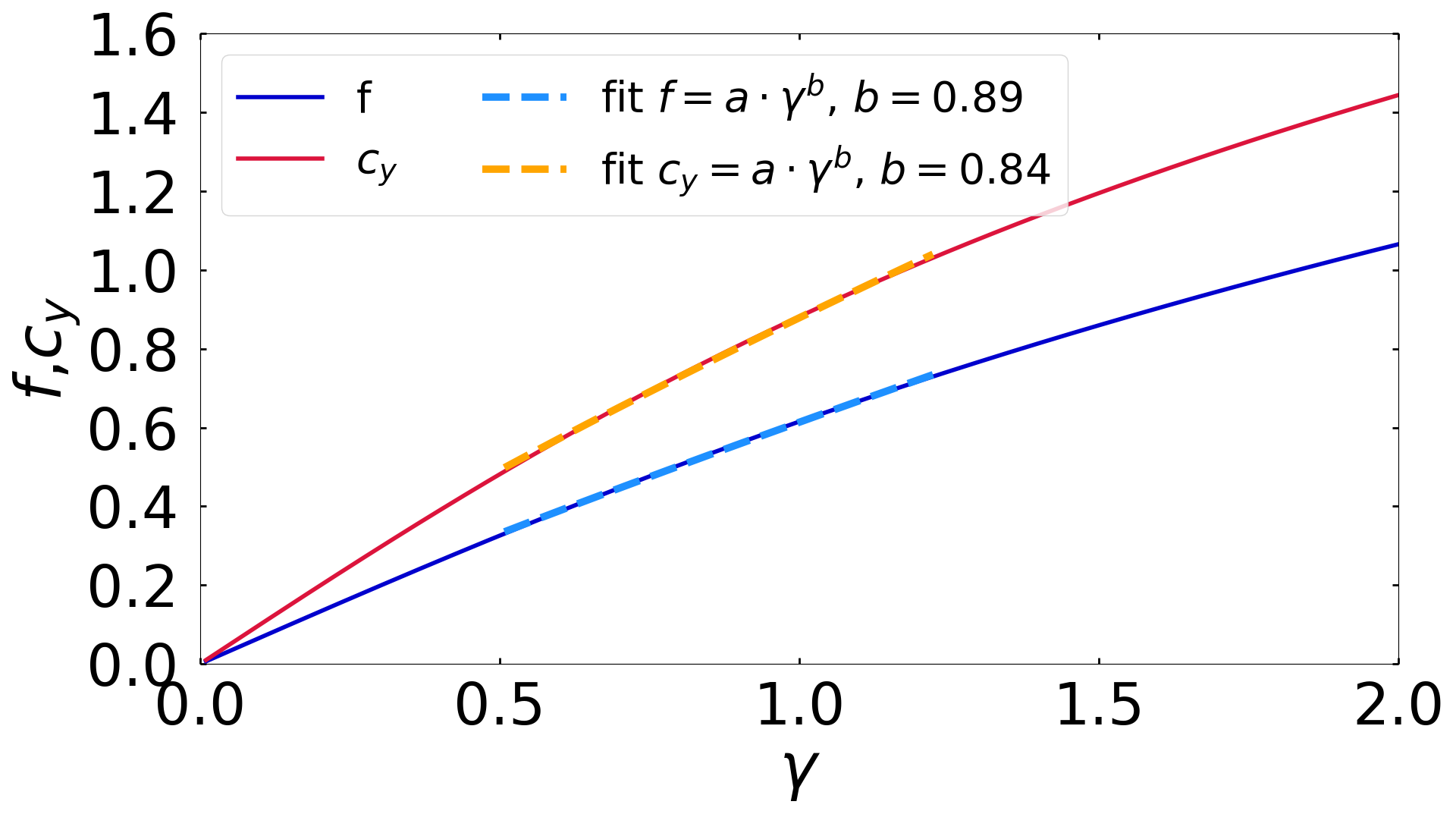}
  \caption{$f(\gamma)$ we find in \cite{bondar2008instantaneous} and $c_y(\gamma)$ as given in \cite{mur2001energy} with the respective power law fits in the $\gamma$-regime defined by the parameters listed in Fig. \ref{New-exponent-for-Shvetsov}.}
  \label{Non-adiabaticWidth}
\end{figure}

These non-adiabatic effects on the intensity dependence of the Rydberg yield can be explained by the width in the distribution of the starting velocity and the ionization phase, which both increase slower with intensity in the non-adiabatic theory than in the adiabatic one. Since this affects the denominator of the Rydberg yield, we end up with a less negative exponent in the power law.
In order to estimate the extent of this effect, we first look at the width $\sigma_{v_\perp} = \sqrt{\omega/(2 c_y)}$ of the transverse velocity distribution for the non-adiabatic case as given in \cite{mur2001energy}. It is 
\begin{equation}
 c_y = \tau_0 = \sinh^{-1}(\gamma) \label{eq:cy}
\end{equation}
which in the adiabatic limit $\gamma \ll 1$ can be approximated by
\begin{equation}
 c_y = \tau_0 \approx \gamma \propto 1/F_0 \propto I^{-0.5}. \label{eq:cy_adiabatic}
\end{equation}
For the non-adiabatic regime used in this paper we fit a power law to eq. \ref{eq:cy} (Fig. \ref{Non-adiabaticWidth}) and obtain $c_y \propto \gamma^{0.84}$ and thus $\sigma_\perp \propto \gamma^{-0.84/2} \propto F^{0.84/2} \propto I^{0.84/4}$. We proceed analogously with the phase width: In \cite{bondar2008instantaneous} the ionization rate is found to have the exponential dependence  $\exp{(\frac{-2 I_p}{\omega}f(\gamma,v_{||},v_{\perp}))}$, so we use $\sigma_{\phi} \propto 1/\sqrt{f}$. In a power law fit to $f(\gamma)$ where we set $v_{||}=0$ and $v_{\perp}=0$ we obtain $f(\gamma) \propto \gamma^{0.89}$ and consequently $\sigma_{\phi} \propto 1/\sqrt{\gamma^{0.89}} \propto F_0^{0.89/2} \propto I^{0.89/4}$ (see Fig. \ref{Non-adiabaticWidth}). 

Consequently, including the non-adiabatic effect both in the velocity and in the phase width we obtain:
\begin{align}
\begin{split}
  &N^*/N \propto \frac{1/\sqrt{I}}{\sigma_\perp \sigma_{\phi}} \\
  &\propto 
  \begin{cases} 1/I^{0.5+0.5}=1/I^{1.0} &\mbox{adiabatic} \\
  1/I^{0.5+0.84/4+0.89/4} = 1/I^{0.933}  &\mbox{non-adiabatic}. \label{eq:adiabatic_nonadiatic_power_law} \end{cases} 
\end{split}
\end{align}

Although this estimate of $b=-0.93$ does not agree perfectly with the power law exponent $b=-0.86$ obtained from the experimental data we got much closer to it. This does not only highlight the relevance of taking account of non-adiabatic effects, but it also shows in what way FTI can be used to investigate the initial conditions at the tunnel exit. In particular, as the discussed effects concern the denominator of the Rydberg yield and thus the total number of tunneled electrons, they are not only relevant for Rydberg related studies but for tunnel ionization in general. For example, the slower growth of the momentum width with intensity when applying non-adiabatic theories as compared to adiabatic theories can also be seen in the data presented in \cite{arissian10,hofmann2016non}.

\begin{figure}
  \centering
  \includegraphics[width=0.45\textwidth]{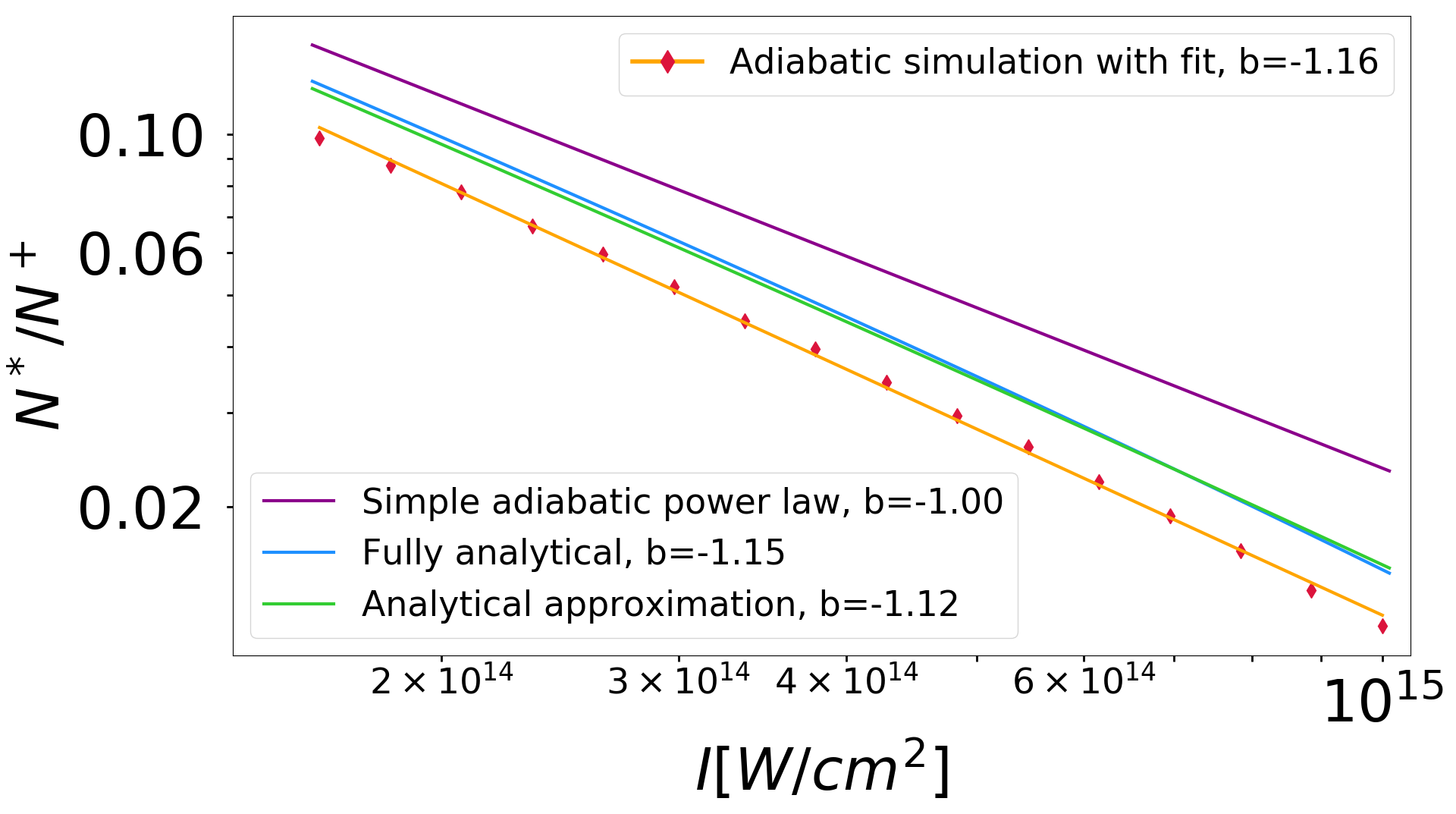}
  \caption{Intensity dependent Rydberg yield at $\lambda=1200 \mathrm{nm}$ (all other parameters are chosen as listed in Fig. \ref{New-exponent-for-Shvetsov}). Purple: The adiabatic power law with $b=-1$ (see eq. \eqref{eq:adiabatic_nonadiatic_power_law}). Red diamonds: Adiabatic CTMC simulation data with power law fit (orange line) to it. 
  Blue: Estimation by solving eq. \eqref{eq:VMaxForResult} and \eqref{eq:phiMaxForResultSimplified} exactly, Green: Approximation given by eq. \eqref{eq:Taylor-approx-final}.}
  \label{lam-dependence}
\end{figure}

For infrared ($\lambda=800 \, \mathrm{nm}$) light the estimation of a power law with exponent $b=-1$ matched the adiabatic simulation results rather well (see Fig. \ref{New-exponent-for-Shvetsov}, $b = -1.02$ for the adiabatic CTMC simulations). 
Since the adiabatic theory is wavelength-independent, we would expect the same scaling 
to hold for larger wavelengths as well - or even better since the system would be more adiabatic. However, the Rydberg yield from adiabatic simulations at $\lambda=1200 \, \mathrm{nm}$ shows a faster drop with intensity which leads to an exponent of $b =-1.16$ in a power law fit (red diamond with orange line in Fig. \ref{lam-dependence}). For larger wavelengths the drop increases even faster with increasing intensity. In the following we derive a theory which explains this effect, thus making predictions about observing this effect in experimental data as well.

As described in \cite{shvetsov2009capture}, we need the maximal initial transverse velocity $v_{\perp,0,max}$ and the range $\Delta \phi$ of ionization phases for estimating the area of initial events in the $v_{\perp,0}-\phi$ plane which end up in a Rydberg state. From \cite{shvetsov2009capture} it becomes clear that including Coulomb effects plays a minor role when dealing with intensity dependence as this effect cancels out in the derivation of $\Delta \phi = |\phi_{latest}-\phi_{earliest}|$  and only shifts the Rydberg area but does not affect its size. 
Hence, we neglect the Coulomb potential in the propagation in the following and `turn on' this potential only at the end of the pulse for the evaluation of eq. \eqref{eq:Rydberg-condition}.

We define the ionization phase of $\phi=0$ to correspond to ionization at the central field maximum, and set the tunnel exit to $x_e=0$. According to the equations of motion in \cite{corkum1993plasma} the position and velocity at a time $\tau$ just after the pulse has passed can be approximated by
\begin{align}
 x(\tau)&\approx\frac{F_0}{\omega^2} \cos{(\phi)}- \frac{F_0}{\omega} \sin{(\phi)} \cdot \tau \label{eq:final_x} \\
 y(\tau)&\approx v_{\perp,0} \cdot \tau \label{eq:final_y}\\
 v_x&=-\frac{F_0}{\omega} \sin{(\phi)} \label{eq:final_vx}\\
 v_\perp &= v_{\perp,0} \label{eq:final_vy},
\end{align} 
where $\tau=T/2$, with $T$ the time span between the zeros of the envelope, and the light is linearly polarized in x-direction.
Note that these equations of motion differ from the ones used in \cite{shvetsov2009capture} by the term $\frac{F_0}{\omega^2} \cos{(\phi)}$ and the $\lambda$-effect in the intensity dependence that we derive arises from this discrepancy. This also explains why the mentioned effect is weakened for longer pulses where the second term in $x(\tau)$ dominates. \\
For the calculation of $v_{\perp,0,max}$ we substitute eq. \eqref{eq:final_x} and \eqref{eq:final_y} in the limit of $E=0$ in eq. \eqref{eq:Rydberg-condition} and set $\phi=0$
\begin{equation}
 E = \frac{v_{\perp,0,max}^2}{2}-\frac{1}{\sqrt{\frac{F_0^2}{\omega^4}+v_{\perp,0,max}^2\cdot \tau^2}}=0. \label{eq:VMaxForResult}
\end{equation}
Analogously, we set $v_{\perp,0}=0$ in the calculation of $\phi_{max}$ in eq. \eqref{eq:Rydberg-condition}, which leads to:
\begin{align}
 \begin{split}
 \frac{1}{2} & \left(\frac{F_0}{\omega}\right)^2 \cdot \sin{(\phi_{max})}^2 \\&=\frac{1}{\frac{F_0}{\omega^2}\cos{(\phi_{max})}-\frac{F_0}{\omega}\sin{(\phi_{max})} \cdot \tau}. \label{eq:phiMaxForResult}
 \end{split}
\end{align}
This expression can be approximated by
\begin{equation}
 \frac{1}{2} \left(\frac{F_0}{\omega}\right)^2 \cdot \phi_{max}^2-\frac{1}{\frac{F_0}{\omega^2}-\frac{F_0}{\omega}\phi_{max} \cdot  \tau} = 0  \label{eq:phiMaxForResultSimplified}
\end{equation}
since $\phi_{max}<0.1$ for the parameters used in this work.
Equations \eqref{eq:VMaxForResult} and \eqref{eq:phiMaxForResultSimplified} can be solved analytically for $v_{\perp,max}$ and $\phi_{max}$, respectively (see Appendix \ref{app:AppendixB} for details).
The corresponding Rydberg yield is estimated as $N^*/N \propto \phi_{max}(F_0,\omega,\tau) \cdot v_{\perp,0,max}(F_0,\omega,\tau) /F_0$ and the intensity dependence at $\lambda=1200 \, \mathrm{nm}$ can be seen in Fig. \ref{lam-dependence} (blue line), a power law fit to which gives an exponent of $b=-1.15$ This analytical derivation matches the simulation data (red diamonds) very well. 
As the lengthy, full analytical solution of \eqref{eq:VMaxForResult} and \eqref{eq:phiMaxForResultSimplified} (see Appendix \ref{app:AppendixB}) does not allow for a deeper understanding of which parameters dominate this wavelength dependence, we also derive an approximation for it in Appendix \ref{app:AppendixC} which yields:
\begin{equation}
 N^*/N \propto \frac{\omega}{F_0^{2} \tau^{2/3} (1+\frac{F_0}{2^{4/3} \cdot \omega^2 \cdot \tau^{2/3}}) }. \label{eq:Taylor-approx-final}
\end{equation}
For the case of $\lambda=1200 \, \mathrm{nm}$, the approximation is plotted in Fig. \ref{lam-dependence} (green line) and a power law fit gives an exponent of $b=-1.12$.
This approximation makes clear that for large wavelengths and small pulse durations the Rydberg yield as a function of intensity is less well described by a power law than for small wavelengths.


In conclusion, we find that including non-adiabatic effects in the distribution of the ionization times and the initial velocity 
leads to a different power law exponent in the intensity dependence of the relative Rydberg yield, resulting in better agreement with experimental data.
As the two mentioned corrections affect the denominator of the Rydberg ratio and thus the total number of electrons that tunneled out of the atom, these insights and approximations can be used beyond studies of Rydberg atoms where one is interested in the intensity dependence of tunnel ionization in a more general context.
Moreover, we find that the power law intensity dependence observed for infrared light breaks down for longer wavelengths.
This correction is based on and highlights the importance of including the offset term $F_0/\omega^2 \cos(\phi)$ in the approximation of the position of an electron that is driven by a laser field.

All in all, these results show new ways to use Rydberg atoms for retrieving information about the tunneling and propagation step in strong field ionization processes.  In particular, measuring Rydberg yield can be used as an independent test for non-adiabatic effects in strong field ionization.  An interesting new twist on Rydberg dynamics is provided by the spatial inhomogeneity of electric fields, such as the one resulting in the vicinity of a nanostructure \cite{ortmann17}.  Under certain conditions, this field inhomogeneity may even lead to chaotic orbits, which should have a significant impact on what fraction of electrons end up in Rydberg states.  

\appendix

\section{} \label{app:AppendixA}
\begin{figure}
  \centering
  \includegraphics[width=0.45\textwidth]{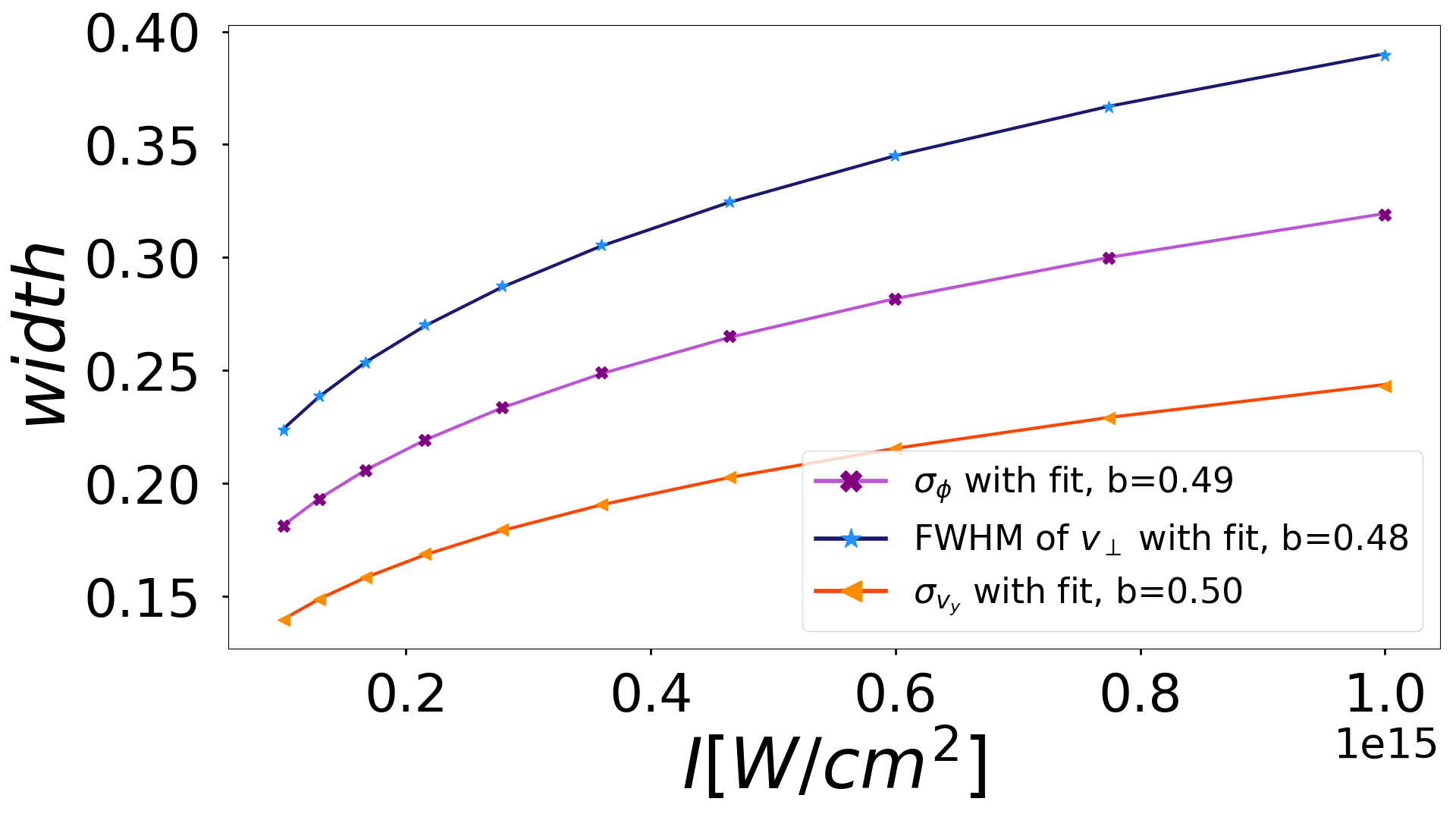}
  \caption{Power law fit ($a F_0^b$ with fitting parameters $a$ and $b$) to the intensity dependence of the standard deviation obtained by Gaussian fits to the histograms of the starting velocities $v_y$ and $v_\phi$ and of the FWHM of the $v_\perp$ histogram for the parameters listed in Fig. \ref{New-exponent-for-Shvetsov}.}
  \label{Field-dependent-width}
\end{figure}

In this section, we show that both the distribution of the initial transverse velocity $v_{\perp,0}$ and of the ionization phase $\phi$ are proportional to $\sqrt{F_0}$ when describing the ionization probability by the adiabatic ADK theory \cite{delone1991energy, ammosov1986tunnel}. 
The histogram of ionization phases qualitatively follows a normal distribution even though formally eq. \ref{eq:ADK-probability} is not Gaussian like. But using the Taylor expansion 
\begin{equation}
 \frac{1}{F(\phi)} = \frac{1}{F_0 \cdot cos(\phi)} \approx \frac{1}{F_0} ( 1 + \frac{1}{2}  \phi^2 + \mathcal O(\phi^4))
\end{equation}
we can rewrite eq. \ref{eq:ADK-probability} as 
\begin{equation}
 P_{approx}=\exp\left(-\frac{\phi^2}{2 \sigma_\phi^2}\right) \quad \text{with } \sigma_\phi = \frac{\sqrt{3\cdot F_0}}{2^{5/4} \cdot I_p^{3/4}}, 
\end{equation}
which makes clear why a normal distribution is a good approximation for it. This also becomes clear from Fig. \ref{Field-dependent-width}, where Gaussian distributions were used to fit the histogram of ionization phases (generated with ADK probability) for various field strengths. The corresponding standard deviations $\sigma_\phi$ are depicted in blue and the power law fit ($a F_0^b$ with fitting parameters $a$ and $b$) to this data gives $\sigma_{\phi} \propto F_0^{0.49}$. 


Also, the dependence of $\sigma_{\perp}$ on the field strength is not as trivial as one might think at first glance. Even though the ionization probability is given by
\begin{equation}
 P(v_{y,0},v_{z,0}) \propto \exp \left(- \frac{v_{y,0}^2+v_{z,0}^2}{2 \sigma_\perp^2}   \right) 
\end{equation}
with
\begin{equation}
 \sigma_\perp^2=\frac{F_0}{2 \cdot (2 I_p)^{1/2}} 
\end{equation}
the probability distribution as a function of $v_{\perp,0}=\sqrt{v_{y,0}^2+v_{z,0}^2}$ has to be transformed into \cite{hofmann2013comparison}
\begin{equation}
 P(v_{\perp,0}) \propto 2 \pi v_{\perp,0} \cdot \exp \left(- \frac{v_{\perp,0}^2}{2 \sigma_\perp^2}   \right) .
\end{equation}
Since this distribution cannot be approximated by a Gaussian, we use the FWHM as a measure for the width. As can be seen from Fig. \ref{Field-dependent-width}, the power law fit gives a field strength dependence on this width of $F_0^{0.48}$ and approximating this by $\sqrt{F_0}$ seems justified. 

\section{} \label{app:AppendixB}
The completely analytical solution to eq. \eqref{eq:phiMaxForResultSimplified} is given by
\begin{equation}
 \phi_{max} = \frac{1+g+\frac{1}{g}}{3 \tau \omega}
\end{equation}
with
\begin{align}
\begin{split}
 g =& F_0^3/\bigg(F_0^9 - 27 F_0^6 \tau^2 \omega^6 ...\\
  &... + 3 \sqrt{-6 F_0^{15} \tau^2 \omega^6 + 81 F_0^{12} \tau^4 \omega^{12}} \bigg)^{1/3}.
\end{split}
\end{align}
Furthermore, equation \eqref{eq:VMaxForResult} is solved exactly by the following expression
\begin{equation}
 v_{\perp,0,max} = \sqrt{\frac{-\frac{F_0^2}{\omega^4}+\frac{F_0^4}{\omega^8 \cdot h}+h}{3 \tau^2}}
\end{equation}
with
\begin{equation}
\begin{split}
 h =& \bigg(-F_0^6 + 6 \omega^6 \Big(9 \tau^4 \omega^6 ...\\
    &...+ \sqrt{-3 F_0^6 \tau^4 + 81 \tau^8 \omega^{12}}\Big)\bigg)^{1/3}/\omega^{4}.
\end{split}
\end{equation}
The Rydberg yield is then estimated by plugging these results into
\begin{equation}
 N^*/N \propto \phi_{max} \cdot v_{\perp,0,max} /F_0.
\end{equation}

\section{} \label{app:AppendixC}
In the following we derive an easy to handle analytical estimation of the Rydberg yield as calculated from equations \eqref{eq:VMaxForResult} and \eqref{eq:phiMaxForResultSimplified}. The idea is to plug the approximate and simpler results $v_{\perp,0,*}=(2/\tau)^{1/3}$ and $\phi_*=-\left(\frac{2}{\tau}\right)^{1/3} \cdot \frac{\omega}{F_0}$ from \cite{shvetsov2009capture} into the Coulomb term of eq. \ref{eq:Rydberg-condition}, which is analogous to solving an equation iteratively. For $v_{\perp,0,max}$ this means
 \begin{align}
 \begin{split}
\frac{v_{\perp,0,max}^2}{2}-\frac{1}{\sqrt{\frac{F_0^2}{\omega^4}+v_{\perp,0,*}^2\cdot \tau^2}} \approx 0 \\
  \Rightarrow v_{\perp,0,max} \approx \frac{\sqrt{2}}{({\frac{F_0^2}{\omega^4}+2^{2/3} \cdot \tau^{4/3})^{1/4}}}. \label{eq:VMaxForResult-analytical}
 \end{split}
\end{align}
And for the phase we obtain
\begin{equation}
 \frac{1}{2} \left(\frac{F_0}{\omega}\right)^2 \cdot \phi^2-\frac{1}{\frac{F_0}{\omega^2}-\frac{F_0}{\omega}\phi_* \cdot  \tau} \approx 0, 
\end{equation}
from which follows
  \begin{align}
 \begin{split}
|\phi_{max}| \approx \frac{\omega/F_0 \cdot \sqrt{2}}{\sqrt{\frac{F_0}{\omega^2}-\frac{F_0}{\omega}(-(\frac{2}{\tau})^{1/3} \cdot \frac{\omega}{F_0}) \tau}} \\
  = \frac{\omega/F_0 \cdot \sqrt{2}}{\sqrt{\frac{F_0}{\omega^2}+2^{1/3}\cdot \tau^{2/3}}}.
 \label{eq:phiMaxForResult-analytical}
 \end{split}
\end{align}

Setting $m=\frac{F_0}{\omega^2}$ and $n=2^{1/3} \cdot \tau^{2/3}$ the Rydberg yield can be expressed as follows:
\begin{align}
 N^*/N &\propto \Sigma^*/\Sigma^+ = \frac{|v_{\perp,0,max}| \cdot |\phi_{max}|}{F_0} \\
  &= \frac{\sqrt{2}}{(m^2+n^2)^{1/4}}  \cdot \frac{\sqrt{2}\cdot \omega/F_0}{(m+n)^{1/2}} \cdot \frac{1}{F_0} \\
  &= \frac{2 \omega}{F_0^{2} \cdot \sqrt{n} \cdot (1+\frac{m^2}{n^2})^{1/4} \cdot \sqrt{n}  \cdot (1+\frac{m}{n})^{1/2}} \\
  &\approx \frac{2 \omega}{F_0^{2} \cdot \sqrt{n} \cdot \sqrt{n}  \cdot (1+\frac{1}{2}\frac{m}{n})} \label{eq:Taylor-approx}\\
  &\propto \frac{\omega}{2^{1/3} \cdot F_0^{2} \cdot \tau^{2/3} (1+\frac{F_0}{2^{4/3} \cdot \omega^2 \cdot \tau^{2/3}}) } \label{eq:Taylor-approx-final}
\end{align}
where in eq. \eqref{eq:Taylor-approx} a Taylor expansion around $m/n\approx0$ is done and the terms with $\mathcal{O}(m^2/n^2)$ are neglected. This expansion to first order  seems reasonable since for the studied parameter regime $m < n$ holds true.

\bibliographystyle{apsrev}
\bibliography{Ryd_lib2}



\end{document}